\documentclass[sigconf]{acmart}

\usepackage{booktabs} 

\usepackage[noend]{algpseudocode}
\usepackage{amsmath}
\usepackage{algorithm}

\makeatletter
\def\BState{\State\hskip-\ALG@thistlm}
\makeatother

\setcopyright{rightsretained}

\copyrightyear{2018} 
\acmYear{2018} 
\setcopyright{acmcopyright}
\acmConference[MM '18]{2018 ACM Multimedia Conference}{October 22--26, 2018}{Seoul, Republic of Korea}
\acmBooktitle{2018 ACM Multimedia Conference (MM '18), October 22--26, 2018, Seoul, Republic of Korea}
\acmPrice{15.00}
\acmDOI{10.1145/3240508.3266443}
\acmISBN{978-1-4503-5665-7/18/10}

\begin{document}
\title{An Iterative Refinement Approach for Social Media Headline Prediction }

\author{Chih-Chung Hsu}
\affiliation{%
  \institution{Department of Management Information Systems, National Pingtung University of Science and Technology (NPUST)}
  \streetaddress{1, Shuefu Road, Neipu}
  \city{Pingtung, Taiwan} 
  \postcode{91201}
}
\email{cchsu@mail.npust.edu.tw}

\author{Chia-Yen Lee}
\affiliation{%
	\institution{Department of Electrical Engineering, National United University (NUU)}
	\streetaddress{2, Lienda}
	\city{Miaoli, Taiwan}
	\postcode{36063}
}
\email{olivelee@nuu.edu.tw}

\author{Ting-Xuan Liao}
\affiliation{%
	\institution{Department of Management Information Systems, NPUST}
}
\email{mdhs213028@mdhs.tc.edu.tw}

\author{Jun-Yi Lee }
\affiliation{%
	\institution{Department of Management Information Systems, NPUST}
}
\email{gaillele85@gmail.com}

\author{Tsai-Yne Hou}
\affiliation{%
	\institution{Department of Management Information Systems, NPUST}
}
\email{tailaeiwa1111@gmail.com}
\author{Ying-Chu Kuo}
\affiliation{%
	\institution{Department of Management Information Systems, NPUST}
}
\email{sun8sophia@gmail.com}
\author{ Jing-Wen Lin}
\affiliation{%
	\institution{Department of Management Information Systems, NPUST}
}
\email{jinwon12081@gmail.com}
\author{ Ching-Yi Hsueh}
\affiliation{%
	\institution{Department of Management Information Systems, NPUST}
}
\email{kub2255@gmail.com}

\author{Zhong-Xuan Zhang}
\affiliation{%
	\institution{Department of Electrical Engineering, NUU}
}
\email{qwert31639@gmail.com}

\author{Hsiang-Chin Chien}
\affiliation{%
	\institution{Department of Electrical Engineering, NUU}
}
\email{shine12100@gmail.com }

\renewcommand{\shortauthors}{C.C. Hsu et al.}

\begin{abstract}
In this study, we propose a novel iterative refinement approach to predict the popularity score of the social media meta-data effectively. With the rapid growth of the social media on the Internet, how to adequately forecast the view count or popularity becomes more important. Conventionally, the ensemble approach such as random forest regression achieves high and stable performance on various prediction tasks. However, most of the regression methods may not precisely predict the extreme high or low values.  To address this issue, we first predict the initial popularity score and retrieve their residues. In order to correctly compensate those extreme values, we adopt an ensemble regressor to compensate the residues to further improve the prediction performance. Comprehensive experiments are conducted to demonstrate the proposed iterative refinement approach outperforms the state-of-the-art regression approach.

\end{abstract}

%
%

\keywords{Random forest, ensemble learning, regression, iterative refinement.}

\maketitle

\section{Introduction}

The popularity prediction of social media becomes more important while the rapid growth of social networks such as Facebook, Flickr, and Pinterest. Once the headline of posts or pictures can be predicted, the advertisement related to the headline can be placed in. We also can synthesize having the high popularity of the post (headline) according to the feature of the headline. Therefore, how to adequately address the headline prediction of the social media remains a significant challenge.

Recently, machine learning approach is widely used in various tasks such as popularity score prediction, object recognition, and time-series signal analysis. For example, a large-scale social media dataset -- Social Media Headline Prediction Dataset (SMHPD) -- is collected in \cite{Wu2017DTCN}. It includes 305,614 metadata, images, and time-zone information. With SMHPD, our goal is to learn the popularity of the posts based on the metadata only without images information due to express prediction purpose. Besides, the content of an image may mismatch to that of the corresponding post, leading to helpless of the prediction task. 

To adequately address the social media headline prediction task, each metadata of SMHPD should be carefully processed. As described in the social media headline prediction task in \cite{Wu2017DTCN}, there are 15 metadata properties in the metadata. It contains a unique picture ID (pid) along with user id (uid). Also, metadata of the picture such as the posted date (date), category it belongs to (cat. and subcat.), concept, path alias for image (alias), whether public to all users (ispublic), media status (status), title, media type (type), all tags (tags), geometric information such as latitude (lat.), longitude (lon.), and geoaccuracy (acc.).

In general, SVR \cite{svr}\cite{smp1}\cite{smp2} and RFR \cite{rfr} show the outstanding performance among the traditional regression models. However, the inputs to SVR needs to pre-process first to avoid the fact that feature with large value will bias the prediction results \cite{svr2}. However, the data types of social media are a significant difference so that the performance of the popularity prediction based on SVR may be suppressed. Random forest regression allowed heterogeneous data such as social media information and achieved high performance. DNNR also achieved excellent performance on various regression tasks \cite{dl}. However, the model selection and training strategy remain a big challenge. Since there are some extreme values of the popularity score of SMHPD, leading to lower performance based on those standard regressors, we propose a novel iterative refinement approach is proposed to resolve this problem in this study.

The main contribution of this paper is two-fold: i) We propose an iterative refinement approach to deal with extreme value regression task, and ii) We carefully treat the social information and analyze feature importance to achieve the best performance.   

The rest of this paper is organized as follows.
Section 2 presents the proposed iterative refinement approach for social media popularity prediction. In Sec. 3, experimental results are demonstrated. Finally, conclusions are drawn in Sec. 4.

\section{The Proposed Iterative Refinement Approach}
\subsection{Data Preprocessing and Analysis}
 Given a metadata of the social information $\mathbf{X}$ with $15$-dimensional vector, the predicted value can be obtained by $y = h(\mathbf{X}, \mathbf{\theta})$ with learned parameters of random forest regression $\mathbf{\theta}$. Since there are some features are in string data types, making the mathematical uncomputable. One of the advanced approaches to transfer string data type to an encoded-vector is word-embedding \cite{wordembedding}. However, the languages of the string descriptions in the metadata are different so that the word-embedding cannot apply in different languages. Instead, we first adopt a simple strategy to make the metadata $\mathbf{X}$ computable and analyze their feature importance. For the features which have high importance, the digitalization of them should be more carefully treated. In order to support this assumption, we adopt RFR to compute the feature importance.
 
First, we adopt the following steps to make the metadata computable as follows:
\begin{itemize}
	\item UID and PID: Converted to integer type.
	\item Cat. and Subcat.: Given a unique integer value for each category or subcategory.
	\item Concept: Converted to the length of the text description.
	\item Alias: Converted to the length of the text description.
	\item IsPublic: Treated false as 0 and true is 1.
	\item Status: Converted the length of the text description.
	\item Title and Tags: Converted to the count of words.
	\item Type: Converted to the length of the text description.
	\item Date: Converted to integer type.
	\item Lat., Acc., and Lon.: Converted to floating-point type.
\end{itemize}

\begin{figure}
	\centering
	\includegraphics[width=0.45\textwidth]{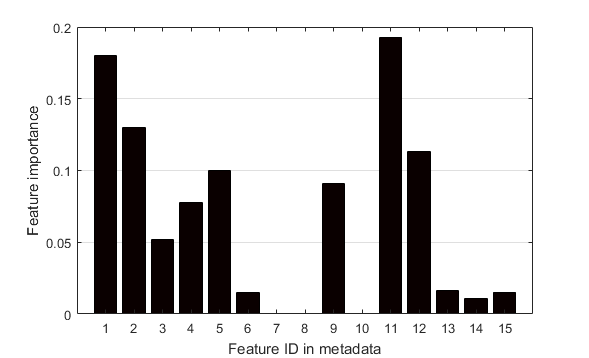}
	\caption{The feature importance analysis of the metadata $\mathbf{x}_c$.}
	\label{fig:fi}	
\end{figure}

As shown in \ref{fig:fi}, it is clear that some of the social information are relatively important than the others. For example, the UID, PID, tags, and date are significantly crucial than others. The social information associated with the identity such as PID and UID is unnecessary to preprocess due to their data is already computable. Consequently, we should pay more attention to the social information which needs to be preprocessed. 

First, we focus on improving the preprocessing strategy of the following five social information: category, subcategory, concept, title, and tags. Since category and subcategory are finite, it is possible to convert this two social information to the unique identity numbering. Toward this end, we first calculate the repeat items and then remove them to find the unique items from the whole social information. Then, each unique item will be given a number to be computable. The pseudo-code is drawn in Algorithm \ref{uniqueid}. Finally, we adopt Algorithm \ref{uniqueid} to obtain the unique ID of category, subcategory, and concept. In order to avoid the problem causing by different languages usage, instead, we convert tags and title to the length of their text description to simplify the learning task. 

\begin{algorithm}
	\caption{Unique ID Converter}\label{uniqueid}
	\begin{algorithmic}[1]
		\Procedure{Find the unique ID}{}
		\State Given a feature $\textbf{X}=[\textbf{x}_1, \textbf{x}_2, ...,\textbf{x}_N]$ and $count=0$
		\State $\textbf{X}_{copy} \gets \textbf{X}$
		\State $\textbf{X}_{copy}=[\textbf{x}_{c_1}, \textbf{x}_{c_2}, ...,\textbf{x}_{c_N}]$
		\For {i=1 to $N$}
			\State	item $\gets$ $\textbf{x}_{c_i}$
			\For {j=1 to $N$}
				\If{item = $\textbf{x}_{c_j}$ and $i$ != $j$}  
				\State Remove($\textbf{x}_{c_j}$) from $\textbf{X}_{copy}$
				\State $count=count+1$
				\EndIf
				
			\EndFor
			
		\EndFor
		\For{$i=0$ to $count$}
		 \State $\textbf{d}_{i} \gets i$
		\EndFor
		
		\For{$i=0$ to $N$}
			\For{$j=0$ to $count$}
				\If{$\textbf{x}_{i}=\textbf{x}_{c_j}$}
				\State	$\textbf{x}_{i} = \textbf{d}_{j}$
				\EndIf
			\EndFor
		\EndFor
		\Return $\textbf{X}$
		\EndProcedure
	\end{algorithmic}
\end{algorithm}

\subsection{Iterative Refinement}
Random forests for regression is based on the growing decision trees with a random vector $f(\mathbf{\theta})$ so that the predictor (i.e., a decision tree) $h(\mathbf{X}, \mathbf{\theta})$ takes on the values as opposed to labels. In general, the prediction task will be defined as a mean-squared error follows:

\begin{equation}
E_{\mathbf{X}, \mathbf{Y}} (\mathbf{Y}-h(\mathbf{X}))^2,
\end{equation}
where $\mathbf{X}$ and $\mathbf{Y}$ are the training samples and labels (popularity scores). It is easy to predict the popularity score for any test sample $\mathbf{X}_t$ based on the learned random forest $h(\mathbf{X}_t, \mathbf{\theta})$ by solving the above equation. In general, the regression methods will adopt the smoothness regularization term to avoid overfitting. However, it will also lead to the smoothing prediction results, causing the extreme value prediction hardly \cite{smooth}. In our case, some of the posts with the extremely high popularity will be more potential to be a headline. It is worth to predict these extreme values correctly. Given the preprocessed social training set $\mathbf{X}_s$, we first obtain the initial prediction value $\mathbf{P}_s$ based on
\begin{equation}
\mathbf{P}_s = h(\mathbf{X}_s, \mathbf{\theta}). 
\end{equation}
Afterward, the residual value between the predicted values and ground truths can be computed as follows:
\begin{equation}
\mathbf{R} = \mathbf{Y} - \mathbf{P}_s.
\end{equation}

\begin{figure*}
	\centering
	\includegraphics[width=0.65\textwidth]{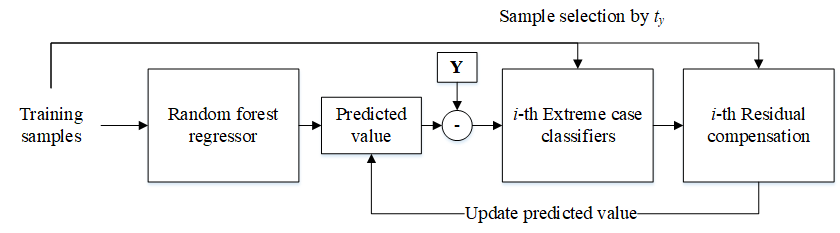}
	\caption{The proposed iterative refinement approach for the social media headline prediction.}
	\label{fig:framework}	
\end{figure*}
By observing the distribution of the residual values, a single one regressor (i.e., random forest) for the social information of SMHPD prediction is unable to exactly predict the higher and lower popularity scores of the posts. It is well-known that a typical regression model tends to fit data distribution in a smoothness way to prevent overfitting issue. 

To effectively compensate the residues of the predicted popularity score $\mathbf{P}_s$, we propose an iterative refinement approach to improve the prediction performance, especially in extreme values compensation. Let the first predicted popularity score be $\mathbf{P}_s$, and its residue is R, the first goal is to learn what samples that would be the extreme high or low popularity scores. Toward this end, it is necessary to learn a classifier that 
\begin{equation}
g(\mathbf{X}_s) = C(\mathbf{X}_s, |\theta_s),
\end{equation}
where $g(\mathbf{X}_s)$ indicates either -1 (non-extreme value) or 1 (extreme value). There are several ways to learn the classifier with the parameters $\theta_s$ such as support vector machine (SVM), random forest classifier (RFC), and AdaBoosting \cite{boosting} classifier. In this study, we adopt AdaBoost as the classifier. In general, the loss function should be 
\begin{equation}
L(\mathbf{X}_s)=\sum_{i=0}^{N}l(C(\mathbf{X}_s, \mathbf{R})),
\end{equation}
where the $l$ is the loss function defined by the learning approach and $L$ is the total loss function. 
Directly solving the above equation is relatively hard because $\mathbf{R}$ is not a binary class. To solve this issue, we predefine a threshold $t_y$ to separate the popularity scores into two groups of larger and regular residues (called $\mathbf{R}_t$). Therefore, the loss function becomes
\begin{equation}
L(\mathbf{X}_s)=\sum_{i=0}^{N}l(C(\mathbf{X}_s, \mathbf{R_t})),
\end{equation}
Intuitively, the larger values in $\mathbf{R}$ indicates a lousy prediction situation, which also implies an extreme value may be presented in $\mathbf{Y}$. 

Let $g_i$ indicate a classifier at $i_{th}$ iteration, the $g_i(\mathbf{X}_s)=1$ should be compensated. Toward this end, we design multi-level regressors to compensate the residues for the prediction each time. Let $h_i$ indicate $i_{th}$ regressor, we need to learn $k$ regressors and classifiers based on the prediction residues $\mathbf{R}_{t_i}$ and the training samples $\mathbf{X}_{R_i}$ with larger residues. Finally, the compensation function at iteration $i$ can be defined as follows:

\begin{equation}
\mathbf{P}_{s_i} = \mathbf{R}_i + \mathbf{P}_{s_{i-1}} 
=h_i(\mathbf{X}_{R_i}, \theta_i) + h_{i-1}(\mathbf{X}_{R_i}, \theta_{i-1}),
\end{equation}
where $\mathbf{X}_{R_i}$ will be $\mathbf{X}_s$ at iteration $0$ and $\mathbf{X}_{R_i} =  [\mathbf{X}_s|g_i(\mathbf{X}_s)=1]$. By controlling the predefined threshold value $t_y$, it is easy to decide the number of the samples to compensate its prediction results. Once we set $t_y=0$, the residual compensation will be performed on all prediction results. The traning process of the proposed iterative refinement approach is illustrated in Fig. \ref{fig:framework}. In test phase, it is quite simple to feed the test sample to the learned RFR and perform the $k$ iterative refinement processes to obtain the final predicted value $\mathbf{P}_{final}$.

\section{Experimental Results}

\subsection{Experimental Settings}
In this experiments, social media headline prediction challenge dataset (SMHPD) containing $305,614$ posts \cite{Wu2017DTCN}\cite{smp1}\cite{smp2} is used to evaluate the performance of the popularity prediction of the proposed method and other state-of-the-art methods. To fairly verify the performance of the proposed method, we partition SMHPD into a $300,000$ training samples and $5,614$ test samples. In the experiments, we have two different partition manners of SMHPD. First, we randomly split SMHPD into training and test sets without considering time-order (Set-A). Second, we follow the instruction in \cite{Wu2017DTCN}\cite{smp1}\cite{smp2} to partition SMHPD into training and test sets in time-order (Set-B). We also download $275,066$ images from Flickr for performance comparison purpose. The unavailable images will be replaced with a black image (i.e., all pixel values in the image are zero). The metrics of the performance comparison are rank correlation (Spearman's Rho)\cite{rc}, Mean Absolute Error (MAE), and Mean Squared Error (MSE), where rank correlation is a nonparametric measure of statistical dependence between the ranking of two variables.

To compare the performance of the popularity prediction, we collect six stat-of-the-art regression methods as follows: 1) Multi-model approach proposed in \cite{hsu2017social}, 2) Standard random forest regressor, 3) SVR with Radial basis kernel, 4) AdaBoosting Regressor \cite{boosting}, 5) Naive Bayer Regressor, and 6) Linear regression.

\subsection{Performance Comparison}

\begin{table}
  \caption{Performance comparison of the different regression methods on test Set-A (Partitioned randomly).}
  \label{tab:exp2}
  \begin{tabular}{cccc}
    \toprule
    Methods &  Rank correlation & MSE & MAE \\
    \midrule
    Naive Bayer Regressor 	& 0.312 & 7.595 & 2.107\\
    SVR 				& 0.351 & 5.411 & 1.846 \\
    Linear Regression 	& 0.423 & 5.068 & 1.785 \\
    AdaBoosting Regression & 0.883 & 1.442 & 0.671 \\
    Random Forest 		& 0.886 & 1.415 & 0.662 \\
    Multi-model Approach \cite{hsu2017social}		& 0.901 & 1.283 & 0.630 \\
    Proposed method 	&$\mathbf{ 0.919}$ & $\mathbf{1.185}$ & $\mathbf{0.593}$ \\
  \bottomrule
\end{tabular}
\end{table}

\begin{table}
	\caption{Performance comparison of the different regression methods on test Set-B (Partitioned by time-order).}
	\label{tab:exp3}
	\begin{tabular}{cccc}
		\toprule
		Methods &  Rank correlation & MSE & MAE \\
		\midrule
		Naive Bayer Regressor 	& 0.417 & 5.196 & 1.814\\
		SVR 				& 0.441 & 4.999 & 1.769 \\
		Linear Regression 	& 0.424 & 5.186 & 1.803 \\
		AdaBoosting Regression & 0.594 & 3.967 & 1.541 \\
		Random Forest 		& 0.886 & 1.418 & 0.663 \\
		Multi-model Approach \cite{hsu2017social}		& 0.846 & 1.838 & 0.748 \\
		Proposed method 	&$\mathbf{ 0.908}$ & $\mathbf{1.193}$ & $\mathbf{0.600}$ \\
		\bottomrule
	\end{tabular}
\end{table}
Table \ref{tab:exp2} shows the prediction performance of the proposed method and six state-of-the-art regression methods for the test Set-A partitioned randomly. With the MSE, MAE, and rank correlation criterion, the proposed iterative refinement approach achieves the best performance, compared to other methods. The performance of the multi-model method \cite{hsu2017social} also achieves good performance. Since the images of the posts are usually noised, the deep neural network in \cite{hsu2017social} may not found enough meaningful information to improve the predicted results. In contrast, the proposed refinement approach concentrates on finding the most useful clues from the metadata of SMHPD to compensate the prediction residual, making the outstanding performance. Compare to our method, other methods cannot achieve promising results due to the highly complex property of the metadata. 

Table \ref{tab:exp3} presents the performance comparison of the proposed method and other methods for test Set-B. However, we note that the overall performance of all regression methods (including our iterative refinement approach) on test Set-B is slightly lower than that of Set-A. A possible reason for the lower performance is that all of the methods do not carefully model any temporal information. 
\subsection{Parameters Selection}
In the proposed method, two critical parameters need to be determined. First one is the predefined threshold value $t_y$ to partition the residues into two groups. A lower threshold value $t_y$ is, the more predicted results will compensate. The second parameter is the number of the iterations of the proposed refinement approach. Intuitively, the more iterations perform, the higher the performance we can achieve. To find the best parameters setting, we conduct two experiments to determine these two parameters. For the selection of parameter $k$, the best performance presented in $k=4$. Note that the experiment is conducted while $t_y=0$. However, it is remarkable that a high performance gain is shown at iteration $2$. In practical, we suggest that the $k$ can be $2$ if it is a time-limited application. Otherwise, $k$ can set to $3-4$ to obtain the best performance. In order to determine another parameter $t_y$, we set the value of $t_y$ to the 80\%, 50\%, 25\%, 12\%,6\%,3\%,1\%,and 0\% of the highest value in the residues and $k=2$. We observed that the best performance presented in $t_y=3\%$. It can be suggested that the $t_y$ can set to a lower value in an adequate resource situation and set to $25\%$ in resource-limited condition. 

In the parameters setting of $k=1$ and $t_y=25\%$, the execution time of the training and testing phases are 1,663.5 and 0.56 seconds respectively. With the parameters setting $k=4$ and $t_y=0\%$, the execution time of the training and testing phases are 9,197.6 and 156.5 seconds respectively.

\section{Conclusions}
In this study, we have proposed an effectively and efficiently iterative refinement approach for social media headline prediction. The main contribution of the proposed method is to address the problem of the extreme value prediction via progressively refining the predicted popularity scores. Since the proposed method is performed on the metadata only, the computational complexity is also relatively low. A comprehensive experiment demonstrated that the proposed method is effective and efficient. 
\begin{acks}
This work was supported in part by the Ministry of Science and Technology of Taiwan under grant
MOST 105-2628-E-224-001-MY3.
\end{acks}

\bibliographystyle{ACM-Reference-Format}
\bibliography{sample-bibliography} 

\end{document}